# Fermi Level Instability as a Way to Tailor Properties of La$_3$Te$_4$

Muhammad Rizwan Khan[1,*], Harshan Reddy Gopidi[1.*], Mateusz Wlazło[1], and Oleksandr I. Malyi[1,#]

[1]Centre of Excellence ENSEMBLE3 Sp. z o. o., Wolczynska Str. 133, 01-919, Warsaw, Poland

*these authors contributed equally to the work
[#]**Email:** oleksandrmalyi@gmail.com

**Abstract:**
Traditionally, the formation of off-stoichiometric compounds is believed to be the growth effect rather than the intrinsic tendency of the system. However, here, using the example of La$_3$Te$_4$, we demonstrate that in n-type gapped metals having a large internal gap between principal band edges and the Fermi level inside of the principal conduction band, Fermi-level instability can develop, resulting in a reduction of formation energy for acceptor defects. Specifically, La vacancies in La$_3$Te$_4$ form spontaneously to produce the acceptor states and remove a fraction of free carriers from the principal conduction band via electron-hole recombination. Such a unique self-doping mechanism allows to stabilize a range of off-stoichiometric La$_{3-x}$Te$_4$ compounds, which have different electronic properties. Moreover, we thus show how controlling synthesis conditions can be used as a knob to reach the target functionality, including controllable metal-to-insulator transition.

Traditional solid-state physics books teach us that vacancy formation costs the energy needed to break the chemical bonds associated with the point defect[1-2]. Such knowledge gives the impression that under equilibrium conditions, the large concentration of defects is only formed at very high temperatures or as the result of the response to intentional doping. This textbook picture follows classical Dalton's paradigm that materials are composed of elements in exact integer ratios (i.e., the law of definite proportions) and is indeed representative of a large number of solids[3]. However, experimentally, it is also known that some compounds can be synthesized with significant off-stoichiometry that can reach tens of atomic percentages and that their properties cannot be described within the stoichiometric composition[4-5]. Moreover, some compounds can be realized in a range of off-stoichiometric compositions having substantially different materials properties[6-9]. Let us consider, for instance, the case of the rare-earth chalcogenides that have been intensively investigated as high-performance thermoelectric materials[6, 10], superconductors[11-12], and topological materials[13]. Experimentally, these compounds can be synthesized with a large deficiency on the cation site (i.e., $Th_{3-x}P_4$). For instance, $La_3Te_4$, a representative example of a rare-earth chalcogenide family, is known to have up to one-ninth of the lanthanum sites vacant[14-15]. The presence of vacancies provides disorder and distortion in the lattice, which does not cause the transition to another crystal structure[6]. What makes these phenomena even more special is that the conductivity and resistivity vs. temperature can be gradually controlled by tuning synthesis conditions stabilizing different $La_{3-x}Te_4$ compounds, even leading to metal-insulator transition[16-17]. Despite such experimental observations, to the best of our knowledge, there is no clear understanding of why such off-stoichiometry can be observed in $La_3Te_4$. Until now, it has been believed that the formation of off-stoichiometric compounds (historically known as Berthollides[18]) is due to the stabilization of the metastable phases and is not linked to the intrinsic tendency of the compound. It is noted that non-stoichiometry is often observed in systems where the ionic species can co-exist in different oxidation states[19]. However, it cannot explain the La deficiency observed in the $La_3Te_4$ system, considering that La and Te usually exist in a single oxidation state. Motivated by this, herein, we try to identify the origin of the non-stoichiometry in $La_3Te_4$ by exploring the defect physics of the compound, explaining also how such non-stoichiometry can affect the electronic properties of the compound. Moreover, despite our focus on $La_3Te_4$, we demonstrate that it belongs to a wide class of materials - gapped metals that can potentially exhibit such non-stoichiometry owing to their unique electronic structure properties.

**$La_3Te_4$ is the gapped metal with the Fermi level in the principal conduction band:** $La_3Te_4$ crystallizes in the $Th_3P_4$-type non-centrosymmetric cubic crystal structure with the space group ($I\bar{4}3d$, No. 220)[15]. Our PBE (i.e., density functional theory calculations using Perdew-Burke-Ernzerhof (PBE)[20] exchange-correlation (XC) functional) band structure calculations show that $La_3Te_4$ is metal (Fig. 1 a,b). However, in contrast to traditional metals (e.g., Ni[21-23]) having the overlap of the principal band edges, $La_3Te_4$ is n-type gapped metal with a Fermi level in the principal conduction band (CB) and a large internal band gap ($E_g^{int}$) of 1.22 eV between the principal band edges. The number of free carriers in the CB is 1e per formula unit (f. u.), and the occupied part of the conduction band $\Delta E_{CB}$ is 0.48 eV. These electronic structure results are consistent with the simplified ionic picture where each La and Te atoms have oxidation states of +3 and -2, respectively. We note that while some gapped metals are simply such due to soft XC functional in naive first-principles calculations (i.e., one that significantly deviates from Koopmans-compliant[24], see discussion in Ref.[25]), even SCAN[26] and HSE06[27] calculations show gapped metal behavior (see Fig. S1a,b). These results thus really suggest that La$_3$Te$_4$ is an example of intrinsic gapped metal[28-29], which is in line with previous theoretical calculations[30] and experimental results showing an increase in resistance of the La$_3$Te$_4$ compound with the temperature[6]. The electronic structure results imply that one should not expect traditional defect physics as in classical metals[31] and instead, there is a possibility of much more complex defect physics similar to insulators and quantum materials[32].

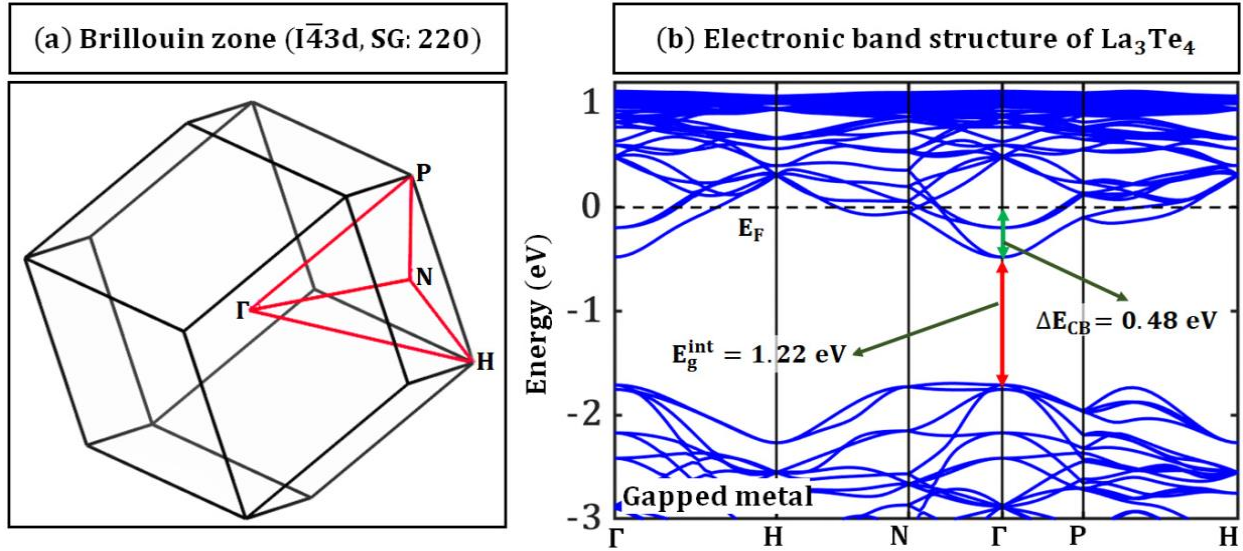

**Figure 1.** (a) The bulk Brillouin zone (BZ) for La$_3$Te$_4$ in cubic (I$\bar{4}$3d, SG: 220) symmetry with several high-symmetry **k**-points Γ (0.0, 0.0, 0.0), N (0.0, 0.0, 0.5), P (0.25, 0.25, 0.25), and H (0.5, 0.5, 0.5). The red lines highlight the high symmetry **k**-path, and (b) the calculated electronic band structure along the high symmetry **k**-paths for La$_3$Te$_4$ using PBE functional, here, the dotted line corresponds to the Fermi level, the internal band gap ($E_g^{int}$) is 1.22 eV, and occupied part of conduction (ΔE$_{CB}$) is 0.48 eV.

**Stability of La$_3$Te$_4$ with respect to competing phases:** To understand the physics behind the formation of non-stoichiometry in La$_3$Te$_4$ compound, first, we calculate the energy convex hull (see methods) for all experimentally known stoichiometric phases of La-Te systems by screening their different binary compounds as well as elemental La and Te. Specifically, we use all known experimentally reported stoichiometric crystal structures available in the Materials Project[33], Open Quantum Materials Database (OQMD)[21-22], and Inorganic Crystal Structure Database (ICSD)[23] listed in Table S1. The results are summarized in Fig. 2, where the formation energy of the compounds as a function of composition is shown. By definition, compounds on the convex hull are stable with respect to decomposition to competing phases (at least for some range of chemical potentials), while if the compound is above the convex hull, it tends to decompose. As can be seen, LaTe (Fm$\bar{3}$m)[34], LaTe$_2$ (Pma2)[35-36], La$_3$Te$_4$ (I$\bar{4}$3d)[15], La$_2$Te$_5$ (P4bm)[34] and LaTe$_3$ (Cmcm)[34] reside on the energy convex hull. All these compounds are widely synthesizable materials with distinct synthesis procedures that are well documented even in ICSD[23]. We note that the stability of the La$_3$Te$_4$ compound is determined by competing with LaTe and LaTe$_2$ compounds, but according to these calculations there is range of chemical potentials under which the compound is thermodynamically stable.

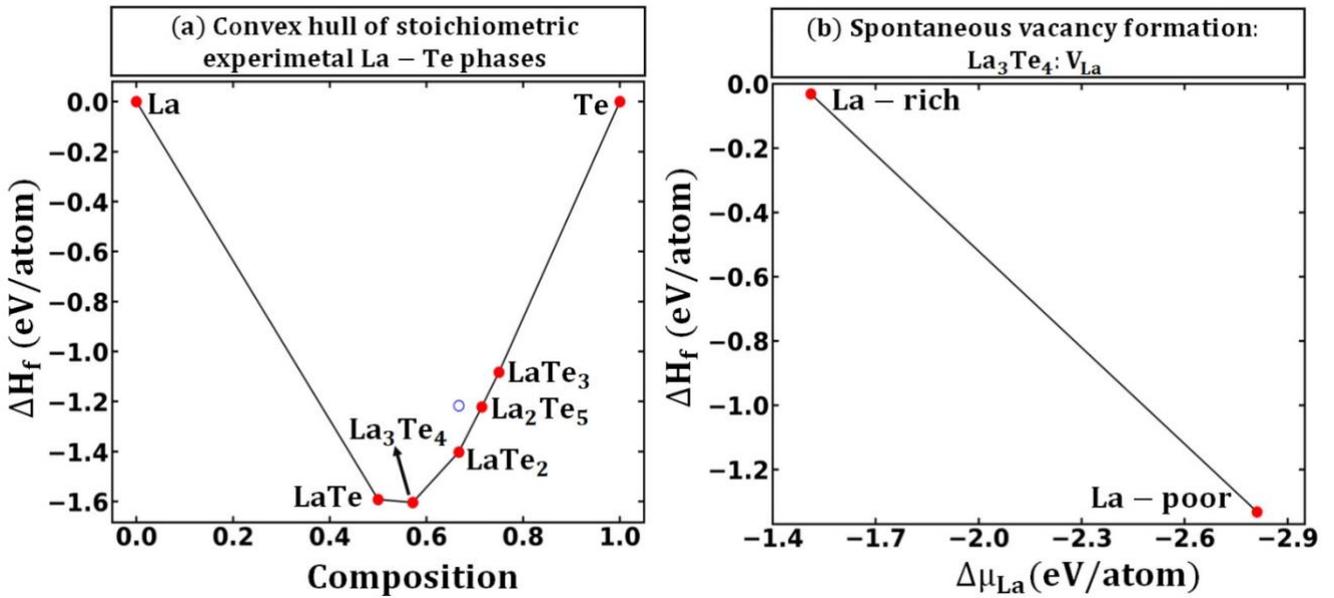

**Figure 2.** (a) Convex hull for the La-Te system, including all known experimentally reported stoichiometric phases. Filled circles represent the compounds with formation energy on the convex hull and empty circle corresponds to other compounds above the convex hull. (b) Defect formation energy for $La_3Te_4$ compound as a function of the chemical potential of La vacancy atom, with respect to La-rich and poor phases.

**La vacancies form spontaneously in $La_3Te_4$:** what makes the case of $La_3Te_4$ special is that in contrast to thermodynamically stable compounds, the formation energy of La vacancies (see methods) is negative for all ranges of chemical potentials under which the compounds are stable with respect to LaTe and $LaTe_2$ compounds (see Fig. 2). Moreover, the defect formation energy can be as low as -1.33 eV for La-poor conditions. We note that all these calculations are performed by density functional theory (DFT) at T=0 K. Hence, it can be concluded that the low defect formation energy is due to the intrinsic tendency of the system and not due to the growth conditions or stabilization of metastable systems. Moreover, these results imply that the convex hull shown in Fig. 2 is likely incomplete. To better understand the physical origin of the results, we reference the acceptor vacancy in a regular insulator, where vacancy formation results in breaking the chemical bonds and unoccupied acceptor states. For the cases of n-type gapped metals, a similar process takes place: (i) vacancy formation still requires the breaking of chemical bonds (which costs energy), leading to (ii) acceptor states. However, for the gapped metal, (iii) electrons in CB can decay to the acceptor level reducing part of the energy needed to break chemical bonds (Fig. 3a, b). Depending on the nature of the state, it is possible that energy lowering due to electron-hole recombination can be greater than the energy gain needed to break the chemical bonds. The latter is the case for $La_3Te_4$, where each La vacancy acts as an acceptor removing 3e per vacancy from the conduction band. This behavior is similar to that recently demonstrated for acceptor defects in gapped metal oxides for $BaNbO_3$[29], $Ag_3Al_{22}O_{34}$[29], and $Ca_6Al_7O_{16}$[29]. Taking these results into account, one can expect that many gapped metals can develop instability with respect to the formation of a large concentration of point defects simply due to the Fermi-level instability described above.

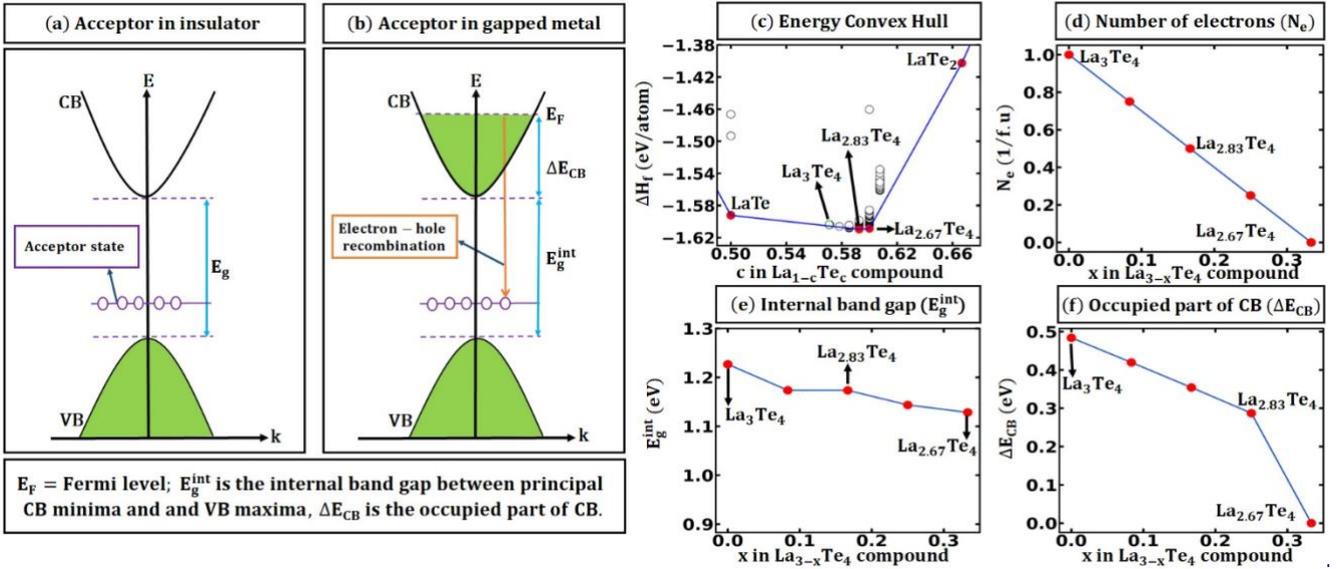

**Figure 3.** Origin of off-stoichiometry in La$_3$Te$_4$. Formation of acceptor vacancy in (a) insulator and (b) n-type gapped metal. Here, E$_F$, ΔE$_{CB}$, and E$_g^{int}$ correspond to the Fermi level, energy range for the occupied part of the conduction band, and internal band gap between the principal valence band maximum and conduction band minimum. (c) Energy convex hull for La-Te system calculated by accounting for non-stoichiometric La$_{3-x}$Te$_4$ phases as a function with composition. Characteristic properties of La$_{3-x}$Te$_4$ as a function of the concentration of La vacancies: (d) number of free carriers per formula unit, (e) internal band gap energy, and (f) energy range for occupied states in the conduction band.

**Spontaneous non-stoichiometry in La$_3$Te$_4$:** Negative defect formation energy implies that La vacancies can form spontaneously until reaching critical defect concentration when the electron-hole recombination cannot further compensate for the energy needed to break chemical bonds. By screening the composition space of off-stoichiometry La$_{3-x}$Te$_4$ systems (see methods), we find that La$_3$Te$_4$ is actually a thermodynamically unstable compound with small energy above convex hull (i.e., 1.6 meV/atom). This can be understood as that convex hull data presented in Fig. 2 are incomplete due to not accounting for non-stoichiometric compounds that indeed can be ground state for specific compositions. We note, however, that such decomposition energy is a relatively small as compared to typical instability reachable experimentally (e.g., 0.1 eV/atom[26, 37]), which thus implies that La$_3$Te$_4$ can still be synthesizable with proper selection of synthesis conditions to eliminate competing phases. We also find that off-stoichiometry on La sites reduces the energy above the convex hull for La$_{3-x}$Te$_4$ compounds. For instance, while La$_3$Te$_4$ is 1.6 meV/atom above the convex hull (when the possibility for the formation of the off-stoichiometry compounds is accounted for), La$_{2.83}$Te$_4$ and La$_{2.67}$Te$_4$ off-stoichiometry systems are on the convex hull. These results are in line with the successful experimental synthesis of La$_{2.82}$Te$_4$, La$_{2.68}$Te$_4$, and La$_{2.67}$Te$_4$[16]. We thus demonstrate that off-stoichiometry of La$_3$Te$_4$ originates Fermi level instability resulting in the formation of point defects removing a fraction of conducting electrons to the formed vacancy acceptor level and reducing the width of the occupied region of the conduction band ΔE$_C$. What makes the off-stoichiometry compounds important is that until reaching 1/9 vacancy concentration (i.e., approximately La$_{2.67}$Te$_4$) all La$_{3-x}$Te$_4$ compounds are n-type gapped metals with Fermi level in the conduction band with tunable free carrier concentration. Here, the number of free electrons per formula unit is defined by the composition weighted sum of oxidation states, assuming La and Te oxidations states to be 3 and -2, respectively. The internal band gap always remains to roughly around 1.16±0.05 eV. When the concentration of La vacancies reaches 1/9, the metal-insulator transition is observed, La$_{2.67}$Te$_4$ is an insulator with the PBE band gap of 1.13 eV. Importantly, increasing La off-stoichiometry beyond 1/9 is thermodynamically unfavorable – for instance, La$_{2.58}$Te$_4$ is 23.7 meV/atom above the convex hull. This is simply because above this critical concentration, La$_{3-x}$Te$_4$ cannot benefit from electron-hole recombination as La$_{2.67}$Te$_4$ does not have any free carriers in the conduction band. All these results thus shed light on not only the origin of the off-stoichiometry in La$_3$Te$_4$ and the origin of metal-insulator transition but which external knobs can be used to tailor properties of La$_{3-x}$Te$_4$. What makes these results even more important is the fact that La$_3$Te$_4$ is a representative example of a wide class of materials, which all potentially can have the same phenomena.

In summary, using the example of La$_3$Te$_4$, we demonstrate a class of materials that can develop spontaneous off-stoichiometry due to Fermi-level instability. Specifically, being a gapped metal with the Fermi level in the principal conduction band and a large internal gap below it, La$_3$Te$_4$ can form La vacancies, resulting in the formation of the acceptor state and removing a fraction of electrons from the conduction band. This eventually leads to the possibility of the formation of a range of off-stoichiometric compounds all having different electronic properties. Moreover, tuning synthesis conditions allow stabilizing a target off-stoichiometric compound with target electronic properties. This work thus shows how spontaneous off-stoichiometry can have an intrinsic origin and be used for a specific application. While we focus on a single compound in this work, the mechanism proposed here represents a full class of materials where the effect can be observed.

**Methods:** We carried out first-principles calculations using Perdew-Burke-Ernzerhof (PBE) functional[20] as implemented in Vienna Ab Initio Simulation Package (VASP)[38-41]. For final static and volume relaxation calculations, the values of cutoff energy for the plane-wave basis were set to be 550 and 500 eV, respectively. For the Brillouin-zone sampling $\mathbf{\Gamma}$-centered Monkhorst-Pack k-grids[42] were used with grid densities of approximately 3,000 and 10,000 per reciprocal atom for volume relaxation and final static calculations. For each system, random atomic displacements within 0.1 Å were applied to avoid trapping at a local minimum. The full optimization of lattice vectors and atomic positions was allowed. The results were analyzed using pymatgen[43].

**Calculations of formation heat of a compound:** For each compound, we calculated the formation heat ($\Delta H_f$) as:

$$\Delta H_f(La_{n_1}Te_{n_2}) = E(La_{n_1}Te_{n_2}) - n_1 E(La) - n_2 E(Te),$$

where, $E(La_{n_1}Te_{n_2})$ is the total ground state energy of the compound, $E(La)$ and $E(Te)$ are the total ground state energies of pure elements in their stable phases, and $n_i$ stands for the number of atoms of the i$^{th}$ element in a single formula unit.

**Calculations of defect formation energy:** The defect formation energy is calculated as

$$\Delta U_f(V_{La}) = E_t^v(La_3Te_4:V_{La}) - E_t(La_3Te_4) + \mu_{La}^{Bulk} + \Delta\mu_{La},$$

where $E_t^v(La_3Te_4:V_{La})$ is the total energy of supercells with La-vacancy (here, we used a 2×2×2 supercell of La$_3$Te$_4$ compound with one La-vacancy), $E_t(La_3Te_4)$ is the total energy of the supercell without vacancy, $\mu_{La}^{Bulk}$ is the chemical potential of La atom in the stable elemental form in hexagonal crystalline symmetry with space group $(P6_3/mmc)$[33] and $\Delta\mu_{La}$ is the chemical potential of La with respect to elemental phase. Here, $\Delta\mu_{La}$ is calculated with respect to LaTe and LaTe$_2$ compounds (see Fig. 2).

**Generation of off-stoichiometric compounds:** To explore the stability of off-stoichiometric La$_{3-x}$Te$_4$ compounds, we used an 84-atom supercell and generated five off-stoichiometric compositions (i.e., supercell containing 1, 2…5 La vacancies). For each composition, we generated at least 40 different randomly selected atomic configurations and performed DFT calculations (with the method described above) for every structurally unique system. In addition to this, for final convex hull calculations we also account for other theoretically known La-Te phases available in Materials Project[33], Open Quantum Materials Database (OQMD)[21-22], and Inorganic Crystal Structure Database (ICSD)[23].

**Acknowledgment:** The authors thank the "ENSEMBLE3 - Centre of Excellence for nanophotonics, advanced materials and novel crystal growth-based technologies" project (GA No. MAB/2020/14) carried out within the International Research Agendas programme of the Foundation for Polish Science co-financed by the European Union under the European Regional Development Fund and the European Union's Horizon 2020 research and innovation programme Teaming for Excellence (GA. No. 857543) for support of this work. We gratefully acknowledge Poland's high-performance computing infrastructure PLGrid (HPC Centers: ACK Cyfronet AGH) for providing computer facilities and support within computational grant no. PLG/2022/015458